\documentclass[prd,preprint,preprintnumbers,amsmath,amssymb]{revtex4}
\usepackage{epsf}
\usepackage{graphicx}
\usepackage{dcolumn}
\usepackage{bm}
\def\be{\begin{equation}}
\def\ee{\end{equation}}

\begin{document}
%
\title[]{Neutrino-impact ionization of atoms in searches for neutrino
magnetic moment}
\author{Konstantin A. Kouzakov}
\affiliation{Department of Nuclear Physics and Quantum
Theory of Collisions, Faculty of Physics, Moscow State University, Moscow 119991, Russia}%
\email{kouzakov@srd.sinp.msu.ru}
\author{Alexander I. Studenikin}
\affiliation{Department of Theoretical Physics, Faculty of
Physics, Moscow State University, Moscow 119991, Russia}%
\email{studenik@srd.sinp.msu.ru}
\author{Mikhail B. Voloshin}
\affiliation{William I. Fine Theoretical Physics Institute,
University of Minnesota, Minneapolis, Minnesota 55455, USA\\
and Institute of Theoretical and Experimental Physics, Moscow,
117218, Russia}%
\email{voloshin@umn.edu}
\begin{abstract}
The ionization of atomic electrons by scattering of neutrinos is
revisited. This process is the one studied in the experimental
searches for a neutrino magnetic moment using germanium detectors.
Current experiments are sensitive to the ionization energy
comparable with the atomic energies, and the effects of the
electron binding should be taken into account. We find that the
so-called stepping approximation to the neutrino-impact ionization
is in fact exact in the semiclassical limit and also that the
deviations from this approximation are very small already for the
lowest bound Coulomb states. We also consider the effects of
electron-electron correlations and argue that the resulting
corrections to the ionization of independent electrons are quite
small. In particular we estimate that in germanium these are at a
one percent level at the energy transfer down to a fraction of
keV. Exact sum rules are also presented as well as  analytical
results for a few lowest hydrogen-like states.

\end{abstract}
\preprint{FTPI-MINN-11/02} \preprint{UMN-TH-2935/11}

\maketitle

\section{Introduction}
The neutrino magnetic moments (NMM) expected in the Standard Model
are very small and proportional to the neutrino masses~\cite{fs}:
$\mu_\nu \approx 3 \times 10^{-19}\, \mu_B \, (m_\nu/1 \, eV)$
with $\mu_B = e/2m$ being the electron Bohr magneton, and $m$ is
the electron mass. Thus any larger value of $\mu_\nu$ can arise
only from physics beyond the Standard Model (a recent review of
this subject can be found in Ref.~\cite{gs}). Current direct
experimental searches~\cite{tx,ge1,ge2} for a magnetic moment of
the electron (anti)neutrinos from reactors have lowered the upper
limit on $\mu_\nu$ down to $\mu_\nu < 3.2 \times 10^{-11} \,
\mu_B$~\cite{ge2}. These ultra low background experiments use
germanium crystal detectors exposed to the neutrino flux from a
reactor and search for scattering events by measuring the energy
$T$ deposited by the neutrino scattering in the detector. The
sensitivity of such a search to NMM crucially depends on lowering
the threshold for the energy transfer $T$, due to the enhancement
of the magnetic scattering relative to the standard electroweak
one at low $T$. Namely, the differential cross section $d
\sigma/dT$ is given by the incoherent sum of the magnetic and the
standard cross section, and for the scattering on free electrons
the NMM contribution is given by the formula~\cite{dn,ve}
\be {d
\sigma_{(\mu)} \over dT }= 4 \pi \, \alpha \,  \mu_\nu^2 \, \left
( {1 \over T} - {1 \over E_\nu } \right ) = \pi \, {\alpha^2 \over
m^2}  \, \left ( {\mu_\nu \over \mu_B} \right )^2 \, \left ( {1
\over T} - {1 \over E_\nu } \right )~ \label{fe} \ee
where $E_\nu$ is the energy of the incident neutrino, and displays
a $1/T$ enhancement at low energy transfer. The standard
electroweak contribution is constant in $T$ at $E_\nu \gg T$:
\be {d
\sigma_{EW} \over dT }= {G_F^2 \, m \over 2 \pi} \left ( 1+ 4 \,
\sin^2 \theta_W + 8 \, \sin^4 \theta_W \right ) \, \left [ 1 + O
\left ( {T \over E_\nu} \right) \right ] \approx 5 \times 10^{-48}
\, {\rm cm^2 / keV}. \label{sew} \ee
In what follows we refer to these two types of contribution to the
scattering as, respectively, the magnetic and the weak.

The current experiments have reached threshold values of $T$ as
low as few keV and are likely to further improve the sensitivity
to low energy deposition in the detector. At low energies however
one can expect a modification of the free-electron formulas
(\ref{fe}) and (\ref{sew}) due to the binding of electrons in the
germanium atoms, where e.g. the energy of the $K_\alpha$ line,
9.89\,keV, indicates that at least some of the atomic binding
energies are comparable to the already relevant to the experiment
values of $T$. Thus a proper treatment of the atomic effects in
neutrino scattering is necessary and important for the analysis of
the current and even more of the future data with a still lower
threshold. Furthermore, there is no known means of independently
calibrating experimentally the response of atomic systems, such as
the germanium, to the scattering due to the interactions relevant
for the neutrino experiments. Therefore one has to rely on a pure
theoretical analysis in interpreting the neutrino data. This problem
had been addressed numerically in the past~\cite{kmsf,fms} and the
interest to the problem was renewed in several recent papers,
which however are ridden by a `trial and error' approach. The
early claim~\cite{wll} of a significant enhancement of the NMM
contribution by the atomic effects has been later
disproved~\cite{mv,wll2} and it was argued~\cite{mv} that the
modification of the formulas (\ref{fe}) and (\ref{sew}) by the
atomic binding effects is insignificant down to very low values of
$T$. It has been subsequently pointed out~\cite{ks} that the
analysis of Ref.~\cite{mv} is generally invalidated in
multi-electron systems, including atoms with $Z > 1$. Furthermore,
the analysis of Ref.~\cite{mv} is also generally invalidated by
singularities of the relevant correlation function in the complex
plane of momentum transfer, so that the claimed behavior of the
cross section at low $T$ applies only in the semiclassical limit,
although, as will be shown here, it gives a very good
approximation to the actual behavior for an electron bound by a
Coulomb potential.

In this paper we revisit the subject of neutrino scattering on
atoms at low energy transfer. We aim at describing this process at
$T$ in the range of few keV and lower, so that the motion of the
electrons is considered as strictly nonrelativistic. Also in this
range the energy of the dominant part of the incident neutrinos
from the reactor is much larger than $T$ and we thus neglect any
terms whose relative value is proportional to $T/E_\nu$ (in
particular, in this range one can neglect the $1/E_\nu$ term in
Eq.(\ref{fe}) in comparison with $1/T$). Furthermore  any recoil
of the germanium atom as a whole results in an energy transfer
less than $2E_\nu^2/M_{Ge}$, which at the typical reactor neutrino
energy is well below the considered here keV range of the energy
transfer. Thus we formally set the mass of the atomic nucleus to
infinity and  neglect any recoil by the atom as a whole.  In
particular, under these conditions the interaction of the neutrino
with the nucleus can be entirely neglected, and only the
scattering on the atomic electrons is to be considered.

We find that in the scattering on realistic atoms, such as
germanium, the so-called stepping approximation works with a very
good accuracy. The stepping approach, introduced in
Ref.~\cite{kmsf} from an interpretation of numerical data, treats
the process as scattering on individual independent electrons
occupying atomic orbitals and suggests that the cross section
follows the free-electron behavior in Eqs.(\ref{fe}) and
(\ref{sew}) down to $T$ equal to the ionization threshold for the
orbital, and that below that energy the electron on the
corresponding orbital is `deactivated' thus producing a sharp
`step' in the dependence of the cross section on $T$. In the
present paper, we consider general relations for the discussed
scattering on atomic systems in Sec.~\ref{genform} and present in
Appendix~\ref{A} sum rules for the theoretical objects involved in
the calculations~\footnote{The sum rules of Appendix~\ref{A}
correct the omissions made in Ref.~\cite{mv}}. In Sec.~\ref{1e} we
prove that for the scattering on individual electrons the stepping
approximation becomes exact in the semiclassical limit, so that
its applicability is improved with the principal number $n$ of the
atomic orbital. We also find by an explicit calculation
(Appendix~\ref{B}) for a hydrogen-like ground state, i.e. at
$n=1$, that the deviation from the stepping behavior is less than
5\% at the worst point, where the energy transfer $T$ is exactly
at the threshold. The accuracy of the approach based on
considering the scattering on individual electrons is limited by
the existence of the electron-electron correlations in the
process. We consider the correction introduced by these
correlations in Sec.~\ref{elcorr} and, in Sec.~\ref{TF}, apply the
derived formula to an estimate of the effect for germanium, using
the Thomas-Fermi model. We find that the correlation correction
grows at smaller $T$ but is still small, of order of a few
percent, for $T$ in the range of a few hundred eV. We thus argue
that the stepping approach describes the scattering cross section
with a sufficient for practical purposes accuracy, and that it can
be applied to the analysis of the present and future data of
searches for NMM with germanium detectors down to the values of
the energy deposition $T \sim 0.3$\,keV.

\section{General formulas for neutrino scattering on atomic electrons}
\label{genform}
In this section we briefly recapitulate the general expressions
and introduce the relevant atomic objects for the neutrino
scattering on atomic electrons. We start with the magnetic process
and then also apply a similar treatment to the standard weak part
of the cross section.

The kinematics of the scattering of a neutrino on atomic electrons
is generally characterized by the components of the four-momentum
transfer, the energy transfer $T$ and the spatial momentum
transfer $\vec q$, from the neutrino to the electrons with two
rotationally invariant variables being $T$ and $q^2= {\vec
q}\,^2$. At small $T$ the electrons can be treated
nonrelativistically both in the initial and the final state, so
that the process is that of scattering of an NMM in the
electromagnetic field $A=(A_0,\vec{A})$ of the electrons:
$A_0(\vec q)=\sqrt{4 \pi \alpha} \, \rho( \vec q)/ {\vec
q}^{\,2}$, $\vec{A}(\vec q)=\sqrt{4 \pi \alpha} \, \vec{j}( \vec
q)/ {\vec q}^{\,2}$, where $\rho(\vec q)$ and $\vec{j}( \vec q)$
are the Fourier transforms of the electron number density and
current density operators, respectively,
\be \rho(\vec q)= \sum_{a=1}^Z \exp(i \vec q
\cdot \vec r_a)~, \label{ne}
\ee
\be \vec{j}(\vec q)= -\frac{i}{2m}\sum_{a=1}^Z \left[\exp(i \vec q
\cdot \vec
r_a)\frac{\partial}{\partial\vec{r}_a}+\frac{\partial}{\partial\vec{r}_a}\exp(i
\vec q \cdot \vec r_a)\right], \label{ce} \ee
and the sums run over the positions $\vec r_a$ of all the $Z$
electrons in the atom.

In this limit the expression for the double differential cross
section is given by~\cite{ks}
\be {d^2 \sigma_{(\mu)} \over dT \, d q^2} = 4 \pi \, \alpha \, {
\mu_\nu^2 \over q^2} \,
\left[\left(1-\frac{T^2}{q^2}\right)S(T,q^2)+\left(1-\frac{q^2}{4E_\nu^2}\right)R(T,q^2)\right]~,
\label{d2s} \ee
where $S(T,q^2)$, also known as the dynamical structure
factor~\cite{vh}, and $R(T,q^2)$ are
\be S(T,q^2)=\sum_n \, \delta (T - E_n+E_0) \, \left | \langle n |
\rho(\vec q) | 0 \rangle \right |^2, \label{dsf} \ee
\be R(T,q^2)=\sum_n \, \delta (T - E_n+E_0) \, \left | \langle n |
j_\perp(\vec q) | 0 \rangle \right |^2, \label{dsfc} \ee
with $j_\perp$ being the $\vec{j}$ component perpendicular to
$\vec{q}$ and parallel to the scattering plane, which is formed by
the incident and final neutrino momenta. The sums in
Eqs.~(\ref{dsf}) and~(\ref{dsfc}) run over all the states $| n
\rangle$ with energies $E_n$ of the electron system, with $|0
\rangle$ being the initial state.

Clearly, the factors $S(T,q^2)$ and $R(T,q^2)$ are related to
respectively the density-density and current-current Green's
functions
\be F(T,q^2)=\sum_n {\left | \langle n | \rho(\vec q) | 0 \rangle
\right |^2 \over T - E_n+E_0 - i \, \epsilon} \,  = \left \langle
0 \left |\rho(- \vec q) \, {1 \over T-H+E_0- i \, \epsilon}\,
\rho(\vec q) \right | 0 \right \rangle~, \label{fdef} \ee
\be L(T,q^2)=\sum_n {\left | \langle n | j_\perp(\vec q) | 0
\rangle \right |^2 \over T - E_n+E_0 - i \, \epsilon} \,  = \left
\langle 0 \left |j_\perp(- \vec q) \, {1 \over T-H+E_0- i \,
\epsilon}\, j_\perp(\vec q) \right | 0 \right \rangle~,
\label{fdefc} \ee
as
\be
S(T,q^2)={1 \over \pi} \, {\rm Im}F(T,q^2)~,
\label{sfrel}
\ee
\be R(T,q^2)={1 \over \pi} \, {\rm Im}L(T,q^2)~, \label{sfrelc}
\ee
with $H$ being the Hamiltonian for the system of electrons. For
small values of $q$, in particular, such that $q\sim T$, only the
lowest-order non-zero terms of the expansion of Eqs.~(\ref{sfrel})
and~(\ref{sfrelc}) in powers of $q^2$ are of relevance (the
so-called dipole approximation). In this case, one has~\cite{ks}
\be R(T,q^2)=\frac{T^2}{q^2}S(T,q^2). \label{da} \ee

Taking into account Eq.~(\ref{da}), the experimentally measured
singe-differential inclusive cross section is, to a good
approximation, given by (see e.g. in Refs.~\cite{mv,ks})
\be
{d \sigma_{(\mu)} \over dT } = 4 \pi \, \alpha \, \mu_\nu^2  \, \int_{T^2}^{4E_\nu^2} \, S(T,q^2)\, {dq^2 \over q^2}~.
\label{d1s}
\ee

The standard electroweak contribution to the cross section can be similarly expressed in terms of the same factor $S(T,q^2)$~\cite{mv} as
\be
{d \sigma_{EW} \over dT } = {G_F^2 \over 4 \pi} \left ( 1+ 4 \, \sin^2 \theta_W + 8 \, \sin^4 \theta_W \right ) \, \int_{T^2}^{4E_\nu^2} \, S(T,q^2) \, dq^2 ~,
\label{d1sw}~,
\ee
where the factor $S(T,q^2)$ is integrated over $q^2$ with a unit weight, rather than $q^{-2}$ as in Eq.(\ref{d1s}).

The kinematical limits for $q^2$ in an actual neutrino scattering
are explicitly indicated in Eqs.(\ref{d1s}) and (\ref{d1sw}). At
large $E_\nu$, typical for the reactor neutrinos, the upper limit
can in fact be extended to infinity, since in the discussed here
nonrelativistic limit the range of momenta $\sim E_\nu$ is
indistinguishable from infinity.  The lower limit can be  shifted
to $q^2=0$, since the contribution of the region of $q^2 < T^2$
can be expressed in terms of the photoelectric cross
section~\cite{mv} and is negligibly small (at the level of below
one percent in the considered range of $T$). For this reason we
henceforth discuss the momentum-transfer integrals in
Eqs.~(\ref{d1s}) and~(\ref{d1sw}) running from $q^2=0$ to
$q^2=\infty$:
\be I_1(T)=\int_0^{\infty} \, S(T,q^2)\, {dq^2 \over q^2}~, \qquad
{\rm and} \qquad I_2(T)=\int_0^{\infty} \, S(T,q^2)\, dq^2~.
\label{defi} \ee

For a free electron, which is initially at rest, the
density-density correlator is the free particle Green's function
\be F_{(FE)}(T,q^2)= \left ( T-{q^2 \over 2m} - i \, \epsilon
\right  )^{-1}~ \label{ff} \ee
so that the dynamical structure factor is given by
$S_{(FE)}(T,q^2)=\delta(T-q^2/2m)$, and the discussed here
integrals are in the free-electron limit as follows:
\be I_1^{(FE)}=\int_0^{\infty} \, S_{(FE)}(T,q^2)\, {dq^2 \over
q^2} = {1 \over T}~, \qquad I_2^{(FE)}=\int_0^{\infty} \,
S_{(FE)}(T,q^2)\, dq^2 = 2 \, m~. \label{intf} \ee
Clearly, these expressions, when used in the formulas (\ref{d1s})
and (\ref{d1sw}), result in the free-electron cross section in
Eqs.~(\ref{fe}) and~(\ref{sew}).

\section{Scattering on one bound electron}
\label{1e}
The binding effects generally deform the density-density Green's
function, so that both the integrals (\ref{defi}) are somewhat
modified. Namely, the binding effects spread the free-electron
$\delta$-peak in the dynamical structure function at $q^2=2 m T$
and also shift it by the scale of characteristic electron momenta
in the bound state. However it turns out that the free electron
expressions are quite robust in the sense that in realistic
systems the modification of the integrals relative to their
free-electron limit, are quite small. As a formal statement, we
will show in the Appendix~\ref{A} that when the function
$F(T,q^2)$ is analytically continued in complex plane of $q^2$ the
free-electron expressions are valid for the integrals over $q^2$
extending from $-\infty$ to $+\infty$, and in the case of the
integral, similar to $I_2$ i.e. with the weight $q^0$, this
property also holds for scattering on multi-electron atomic
systems, while for that with the weight $q^{-2}$ it generally
holds only for the scattering on one electron, or on independent
electrons. Clearly the latter integrals over the full axis of
$q^2$ differ from those of physical interest in Eq.(\ref{defi}) by
the contribution of negative $q^2$, which although numerically
small even at low $T$, still makes the scattering on bound
electrons different from that on free electrons.

In this section we consider the scattering on just one electron.
The Hamiltonian for the electron has the form $H=p^2/2m + V(r)$, and
the density-density Green's function from Eq.(\ref{fdef}) can be written as
\begin{eqnarray}
F(T,q^2)&=& \left \langle 0 \left |\, e^{-i \vec q \cdot \vec r} \, \left [ T -H(\vec p, \vec r) + E_0 \right ]^{-1} \,  e^{i \vec q  \cdot \vec r}\, \right | 0 \right \rangle \nonumber \\
&=&\left \langle 0 \left | \, \left [ T -H(\vec p + \vec q, \vec
r) + E_0  \right ]^{-1} \,  \right | 0 \right \rangle \nonumber\\
&=&
 \left \langle 0 \left | \, \left [ T -{\vec q^{\,2} \over 2 m}- {\vec p \cdot \vec q \over m}  - H(\vec p, \vec r) + E_0  \right ]^{-1} \,  \right | 0 \right \rangle \, ,
\label{f1}
\end{eqnarray}
where the infinitesimal shift $T \to T - i \epsilon$ is implied.

Clearly, a nontrivial behavior of the latter expression in
Eq.(\ref{f1}) is generated by the presence of the operator $(\vec
p \cdot \vec q)$ in the denominator, and the fact that it does not
commute with the Hamiltonian $H$. Thus an analytical calculation
of the Green's function as well as the dynamical structure factor
is feasible in only few specific problems. In Appendix~\ref{B} we
present such a calculation for ionization from the $1s$, $2s$ and
$2p$ hydrogen-like states. In particular, we find that the deviation
of the discussed integrals (\ref{defi}) from their free values are
very small: the largest deviation is exactly at the ionization
threshold, where, for instance, each of the $1s$ integrals is
equal to the free-electron value multiplied by the factor $(1-7 \,
e^{-4}/3) \approx 0.957$~\footnote{It can be also noted that both
integrals are modified in exactly the same proportion, so that
their ratio is not affected at any $T$: $I_2(T)/I_1(T)= 2 m \, T$.
We find however that this exact proportionality is specific for
the ionization from the ground state in the Coulomb potential.}.

The problem of calculating the integrals (\ref{defi}) however can
be solved in the semiclassical limit, where one can neglect the
noncommutativity of the momentum ${\vec p}$ with the Hamiltonian,
and rather treat this operator as a number vector. Taking also
into account that $(H-E_0) \, |0 \rangle =0$, one can then readily
average the latter expression in Eq.(\ref{f1}) over the directions
of ${\vec q}$ and find the formula for the dynamical structure
factor: 
\be S(T,q^2)={m \over 2 \, p \, q} \, \left [ \theta \left
( T- {q^2 \over 2m}+{p \,q \over m} \right) - \theta \left ( T-
{q^2 \over 2m}- {p \, q \over m} \right) \right ]~, 
\label{scls}
\ee 
where $p= |{\vec p}|$ and $\theta$ is the standard Heaviside
step function. The expression in Eq.(\ref{scls}) is nonzero only
in the range of $q$ satisfying the condition $-p\, q/m < T -
q^2/2m < p \,q/m$, i.e. between the (positive) roots of the
binomials in the arguments of the step functions:
$q_{min}=\sqrt{2m \, T + p^2} - p$ and $q_{max}=\sqrt{2m \, T +
p^2} + p$. One can notice that the previously mentioned `spread
and shift' of the peak in the dynamical structure function in this
limit corresponds to a flat pedestal between $q_{min}$ and
$q_{max}$. The calculation of the integrals (\ref{defi}) with the
expression (\ref{scls}) is straightforward, and yields the
free-electron expressions (\ref{intf}) for the discussed here
integrals in the semiclassical (WKB) limit~\footnote{The
appearance of the free-electron expressions here is not
surprising, since the equation (\ref{scls}) can be also viewed as
the one for scattering on an electron boosted to the momentum
$p$}:
\be I_1^{(WKB)}={1 \over T}~,\qquad I_2^{(WKB)}=2 \, m~.
\label{wkbi} \ee
The difference from the pure free-electron case
however is in the range of the energy transfer $T$. Namely, the
expressions (\ref{wkbi}) are applicable in this case only above
the ionization threshold, i.e. at $T \ge |E_0|$. Below the threshold
the electron becomes `inactive'.

It is instructive to point out that the validity of the  result in
Eq.(\ref{wkbi}) is based on the semiclassical approximation and is not directly
related to the value of the energy $T$. In particular, for a Coulomb interaction
the WKB approximation is applicable at energy near the threshold~\cite{ll}. For
$T$ exactly at the threshold, $T=-E_0$, the criterion for applicability of the
semiclassical approach in terms of the force $F=|\vec F| = |[\vec p, H]|$ acting
on the electron and the momentum $p$ of the electron is that~\cite{ll} the ratio
of the characteristic values $m F/p^3$ is small. For the excitation of a state
with the principal number $n$  this ratio behaves parametrically as $1/n$~\footnote{Indeed one has $|F|=\alpha/r^2 \sim m^2 \, \alpha^3 \, n^4$ and $p \sim m \, \alpha/n$, so that $m |F|/p^3 \sim 1/n$.}. Thus
the applicability of a semiclassical treatment of the ionization near the
threshold improves for initial states with large $n$. As previously mentioned,
the modification of the integrals (\ref{defi}) by the binding is already less
than 5\% for $n=1$, so that we fully expect this deviation to be  smaller for
the higher states, and even smaller at larger values of $T$ above the threshold
due to the approach to the free-electron behavior at $T \gg E_0$.

We believe that the latter conclusion explains the so-called
stepping behavior observed empirically~\cite{kmsf} in the results
of numerical calculations. Namely the calculated cross section $d
\sigma/dT$ for ionization of an electron from an atomic orbital
follows the free-electron dependence on $T$ all the way down to
the threshold for the corresponding orbital with a very small, at
most a few percent, deviation. This observation lead the authors
of Ref.~\cite{kmsf} to suggest the stepping approximation for the
ratio of the atomic cross section (per target electron) to the
free-electron one:
\be f(T) \equiv { d \sigma / dT \over (d \sigma/dT)_{FE}} = {1
\over Z} \, \sum_i \, n_i \, \theta(T - |E_i|)~, \label{step} \ee
where the sum runs over the atomic orbitals with the binding
energies $E_i$ and the filling numbers $n_i$. Clearly, the factor
$f(T)$ simply counts the fraction of `active' electrons at the
energy $T$, i.e. those for which the ionization is kinematically
possible. For this reason we refer to $f(t)$ as an {\it activation
factor}. We conclude here that the stepping approximation is
indeed justified with a high accuracy in the approximation of the
scattering on independent electrons, i.e. if one neglects the
two-electron correlations induced by the interference of terms in
the operator $\rho(\vec q)$ in Eq.(\ref{ne}) corresponding to
different electrons. In the next section we estimate the effect of
such an interference and find that the resulting corrections are
small, at least in atoms with large $Z$, such as the germanium.

\section{Two-electron correlation}
\label{elcorr}
In this section we discuss the correction arising from a correlation between two electrons. We consider the energy $T$ and hence the relevant momentum transfer $q$ as large in comparison with the atomic scale. In this way we estimate the relevant parameter for the significance of the correlation effect.

We start with consider an isolated system of two electrons interacting among themselves through the Coulomb potential $V(r)$. The Hamiltonian for this system thus has the form
\be
H={P^2 \over 4m} + {p^2 \over m} + V(r)~,
\label{h2}
\ee
where $\vec P=-i \partial / \partial \vec R$ and $\vec p=-i \partial / \partial \vec r$ are, as usual, the momenta conjugate to respectively the center of mass coordinate $\vec R$ and the relative coordinate $\vec r$.

The spatial part of the wave function of the system factorizes into the product $\phi(\vec R) \, \psi(\vec r)$ with $\phi(\vec R)$ while the spin part will be considered later. We consider here the system at rest, i.e. $\phi(R) = {\rm const}$, since a boost to a momentum $P$ does not change the cross section.

The density-density Green's function then takes the form
\be
F(T,q^2)= \sum_n {\left \langle 0 \left | e^{i \vec q \vec r /2} +  e^{- i \vec q \vec r /2} \right | n \right \rangle \, \left \langle n \left | e^{i \vec q \vec r /2} +  e^{- i \vec q \vec r /2} \right | 0 \right \rangle \over T-{q^2 \over 4 m} - E_n + E_0}~,
\label{f2}
\ee
where the states $|0 \rangle,~|n \rangle$ and the energies $E_0,~E_n$ refer to the relative motion in the system with $|0 \rangle$ standing for the initial state. Clearly, it is implied in Eq.(\ref{f2}) that the corresponding matrix elements for the (trivial) dynamics of the system as a whole are already taken, which results in replacing in the energy denominator the excitation energy $T$ by its value corrected for the recoil of the system as a whole: $T \to T - q^2/4m$.

The cross terms between $\exp(i \vec q \cdot \vec r/2)$ and $\exp(-i \vec q \cdot \vec r/2)$ in the expression (\ref{f2}) result in the previously discussed one-particle Green's function
\be
F_1(T,q^2)=2 \, \left \langle 0 \left | \left ( T - {q^2 \over 2m} - {\vec p \cdot \vec q \over m} - H + E_0 \right )^{-1} \right | 0 \right \rangle~,
\label{f21}
\ee
where the overall factor of 2 arises from the two identical (after averaging over the direction of $\vec q$) cross terms,  and physically is corresponding to the presence of two particles in the system.

The discussed here contribution of the two-electron correlation arises from the diagonal terms, whose contribution is given by
\be
F_c(T,q^2)= 2 \, \eta \, \sum_n {\left \langle 0 \left | e^{i \vec q \vec r /2}  \right | n \right \rangle \, \left \langle n \left | e^{i \vec q \vec r /2}  \right | 0 \right \rangle \over T-{q^2 \over 4 m} - E_n + E_0}~,
\label{fc}
\ee
where, again, the two terms give the same contribution after the averaging over the direction of $\vec q$, which is accounted for by the factor of 2 in the latter expression. The factor $\eta$ is the symmetry factor for the spin part of the two-electron system: $\eta=-1$ for the spin-singlet state of the pair and $\eta=+1$ for the spin-triplet. The appearance of this factor can be explained as follows.  The discussed correlation arises from the situation where an excitation of one electron by the operator $\rho$ produces the same spatial wave function as an excitation of another one. In order for the wave functions to be identical the spin variables of the two electrons should also be switched, which operation results in the factor $\eta$. Clearly, no such factor arises in the one-particle term (\ref{f21}) since the spin of both electrons simply `goes through'. It can be also mentioned that, naturally, the symmetry of the spatial wave function $\psi(\vec r)$ is opposite to $\eta$.

One can notice that unlike in the one-particle contribution (Eq.(\ref{f21})) where the momentum $\vec q$ flows in and out of the system, the correlation contribution in Eq.(\ref{fc}) corresponds to the net momentum $\vec q$ flowing into the system. Clearly, for non-interacting particles such contribution would vanish and the whole correlation effect arises only due to the interaction between the electrons, which interaction absorbs the momentum transfer. The term (\ref{fc}) can be graphically represented as shown in Fig.1, where the system lines correspond to the propagation of the system in the potential $V$ with the outer legs corresponding to the wave functions of the initial state in the momentum space $ \langle \psi(\vec p\,')|$ and $|\psi(\vec p) \rangle$ and the line between the action of the operators $\exp(i \vec q \cdot \vec r/2)$ corresponding to the Green's function. One can write in terms of these objects the expression for $F_c(T,q^2)$ as
\be
F_c(T,q^2)= 2 \int \, {d^3 p' \over (2 \pi)^3} \, \int \, {d^3 p \over (2\pi)^3 } \, \psi^*(\vec p\,') \, G(E; \vec p\,'- {\vec q \over 2}, \vec p + {\vec q  \over 2}) \, \psi(\vec p)~,
\label{fcp}
\ee
where $G(E, \vec k\,', \vec k)$ is the Green's function $(E-H)^{-1}$ in the momentum representation at the energy $E=T+E_0-q^2/4m$.
\\
\begin{figure}[ht]
\begin{center}
 \leavevmode
    \epsfxsize=6cm
    \epsfbox{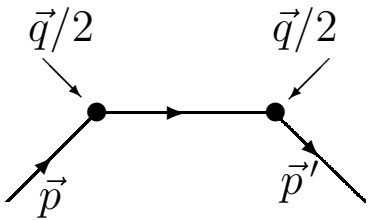}
    \caption{Graphical representation of the two-electron correlation. The external legs correspond to the momentum-space wave function and the propagator is the Green's function in the potential $V$.}
\end{center}
\end{figure}
We shall consider separately the effect of the interaction in the wave functions and in the Green's function. For the zeroth order Green's function
\be
G_0(E, \vec k\,', \vec k)={(2 \pi)^3 \, \delta^{(3)}(\vec k\,' - \vec k) \over E-k^2/m}
\label{g0}
\ee
and the exact wave functions one finds
\be
F_c^{(0)}(T,q^2)=2 \, \eta \, \int \, {d^3 p \over (2\pi)^3 } \, {\psi^*(\vec p +\vec q) \, \psi(\vec p) \over T-{q^2 \over 2 m} - {( \vec p \cdot \vec q) \over m } - {p^2 \over m} +E_0}~.
\label{fc0}
\ee

Let us consider now $q$ as a large parameter in comparison with the characteristic momenta $p_0$ in $\psi(\vec p)$, beyond which the wave function falls off. At such values of $q$ the product
$\psi^*(\vec p + \vec q) \, \psi(\vec p)$ carries a suppression in only one of the factors in two regions of $\vec p$: one where $p \sim p_0$ and the other where $|\vec p + \vec q| \sim p_0$. Clearly, by shifting the integration variable $\vec p + \vec q \to \vec p$ one can readily see that both integration regions the contribution of the latter region is the same as of the first one, so that one evaluates the integral in Eq.(\ref{fc0}) by considering only the  contribution of the region $p \ll q$ and taking it with a factor of two. Then the leading at large $q$ expression for the function in Eq.(\ref{fc0}) takes the form
\be
F_c^{(0)}(T,q^2) =  4 \, \eta \, { \psi^*(\vec q) \over T-{q^2 \over 2 m} } \, \int \, {d^3 p \over (2\pi)^3 }  \, \psi(\vec p)= 4 \, \eta \, { \psi^*(\vec q) \, \psi(0) \over T-{q^2 \over 2 m} }~,
\label{fc0q}
\ee
The appearance of the wave function at the origin $\vec r=0$, $\psi(0)$, in this expression implies that at large $q$ the considered contribution to the correlation arises only for an $S$-wave relative motion within the electron pair, which thus have to be a spin-singlet, and therefore $\eta=-1$.

In fact for an for an $S$-wave motion the momentum-space wave function $\psi(\vec q)$ can be also  expressed  at large $q$ in terms of the position-space wave function at the origin $\psi(0)$. Indeed, in the $S$-wave the wave function is a function of $r$: $\psi(r)$. At small $r$ the Coulomb repulsion between the electrons dominates over all other interactions and the Schr\"odinger equation for $\psi(r)$ reads
\be
-{1 \over m}\, \psi'' -{2 \over m \ r} \, \psi'(r) + {\alpha \over r} \, \psi(r) = E \, \psi(r)~.
\ee
By requiring the two singular at $r \to 0$ as $1/r$ terms to match in this expression, one finds that the derivative of $\psi(r)$ over $r$ at the origin is expressed in terms of $\psi(0)$:
\be
\psi'(0)= {m \, \alpha \over 2} \, \psi(0)~.
\label{der0}
\ee
A finite derivative over $r$ implies that the gradient $\vec \nabla \psi(r)= \vec r \, \psi'(r) /r$ is singular at $r=0$, so that the asymptotic at large $|\vec q|$ behavior of the momentum space wave function is proportional to $1/|\vec q|^4$ with the coefficient determined by $\psi'(0)$, which in turn is determined, according to Eq.(\ref{der0}), by $\psi(0)$:
\be
\left . \psi(\vec q) \right |_{|\vec q| \to \infty} = - {4 \pi \, m \, \alpha \, \psi(0) \over |\vec q|^4}~.
\label{asq}
\ee
Using this relation in Eq.(\ref{fc0q}) one finds in the large $q^2$ limit
\be
F_c^{(0)}(T,q^2) = 16 \, \pi \, \alpha  \,  { m \, |\psi(0)|^2 \over q^4} \, {1 \over T-{q^2 \over 2 m} }~.
\label{fc0f}
\ee

The latter expression is manifestly proportional to first power of the interaction between the electrons. Therefore a similar contribution can arise  from the
first order in the expansion of the Green's function in the interaction potential $V$. In this order one finds for the discussed correlation part of $F(T,q^2)$:
\be
F_c^{(1)}(T,q^2) =  2 \eta \, \int \, {d^3 p' \over (2 \pi)^3} \, \int \, {d^3 p \over (2\pi)^3 } \,{ \psi^*(\vec p\,') \, V(-\vec q+\vec p\,'-\vec p)  \, \psi(\vec p) \over \left [ T-{q^2 \over 2 m} + {( \vec p\,' \cdot \vec q) \over m } - {p'^2 \over m} +E_0 \right ] \, \left [ T-{q^2 \over 2 m} - {( \vec p \cdot \vec q) \over m } - {p^2 \over m} +E_0 \right ]}~,
\label{fc1}
\ee
where $V(\vec k)$ is the Fourier transform of the potential, so that for the Coulomb repulsion between the electrons
\be
V(\vec k)= {4 \pi \alpha \over k^2}.
\label{vk}
\ee

Considering as before the limit of large $q$ and thus neglecting $p$ and $p'$ in comparison with $q$, one readily finds that the result is again proportional to $|\psi(0)|^2$, so that the effect remains only in the $S$-wave (and hence $\eta=-1$):
\be
\left. F_c^{(1)}(T,q^2) \right |_{mT,q^2 \gg p^2, p'^2} = -8 \pi \, \alpha { |\psi(0)|^2 \over q^2 \, \left ( T-{q^2 \over 2 m} \right )^2 }
\label{fc11}
\ee

Collecting the formulas (\ref{fc0f}) and (\ref{fc11}) together one finds the estimate of the two-electron correlation part of $F(T,q^2)$ in the limit of large $T$ and $q^2$:
\be
F_c(T,q^2) = 8 \pi \, \alpha \, |\psi(0)|^2 \,  \left [ {2 \, m \over q^4 \, \left ( T-{q^2 \over 2 m} \right )} - {1 \over q^2 \, \left ( T-{q^2 \over 2 m} \right )^2 } \right ] ~.
\label{fcff}
\ee
The corresponding two-electron correlation correction to the integrals for the neutrino scattering cross section, is then calculated by shifting $T \to T- i \epsilon$ and considering only the contribution of the singularity at $q^2 = 2m \, T$:
\be
{1 \over \pi} \, \int \, {\rm Im} F_c(T,q^2) \, {d q^2 \over q^2}= - 4 \pi \, \alpha \, {|\psi(0)|^2 \over m \, T^3}~, ~~~~~ {1 \over \pi} \, \int \, {\rm Im}F_c(T,q^2) \, d q^2 = 0~.
\label{dic}
\ee
Notice, that the two-electron contribution to the integral relevant for the standard electroweak scattering vanishes in the discussed approximation due to a cancellation of the two terms in Eq.(\ref{fcff}).

The discussed here calculation shows, as expected, that at large $q$ the two-electron correlation arises only when the electrons are separated by a short distance. For this reason one can relax the assumption, we made in the beginning of this section, that the system of two electrons is in a free motion. Indeed the same result would apply in the situation, where the pair as a whole moves in a potential that is sufficiently smooth so that the `tidal force' interaction with the rest of the atomic system does not overcome the Coulomb singularity of the repulsion between the electrons at distances of order $1/q$.

\section{Scattering on atomic electrons in germanium}
\label{TF}
In considering the neutrino scattering on actual atoms one needs
to evaluate the dependence of the number of active electrons on
$T$ and generally also evaluate the effect of the two-electron
correlations. The energies of the inner $K,~L$ and $M$ orbitals in
the germanium atom are well known (see e.g. Ref.~\cite{fms} and
references therein) and provide the necessary data for a
description of the neutrino scattering by the stepping formula
(\ref{step}) down to the values of the energy transfer $T$ in the
range of the binding of the $M$ electrons, i.e. at $T > |E_M|
\approx 0.18$\,keV. The corresponding steps in the activation
factor are shown in Fig.~2. It can be mentioned that if one
applies formulas of the Appendix~\ref{B} to the onset of the $K$
shell step, i.e. just above 10.9\,keV, the difference from the
shown in the plot step function would be practically invisible in
the scale of Fig.~2.

Our goal in this section is to estimate the effect of the two-electron correlations in the scattering on germanium. We shall estimate this effect by considering the atomic number $Z$ as a large parameter and using the Thomas-Fermi model, which, in spite of its known shortcomings, appears to be appropriate for evaluating average bulk properties of atomic electrons at large $Z$, such as in the problem at hand.

In the Thomas-Fermi model (see e.g. Ref.~\cite{ll}) the atomic electrons are described as a degenerate free electron gas in a master potential $\phi(r)$ filling the momentum space up to the zero Fermi energy, i.e. up to the momentum $p_0(r)$ such that $p_0^2/2 m - e \phi=0$. The electron density $n(r)=p_0^3/(3 \pi^2)$ then determines the potential $\phi(r)$ from the usual Gauss equation. In the discussed picture at an energy transfer $T$ the ionization is possible only for the electrons whose energies in the potential are above $-T$, i.e. with momenta above $p_T(r)$ with $p_T^2/2m - e \phi = -T$. The electrons with lower energy are inactive. Calculating the density of the inactive electrons as $p_T^3/(3 \pi^2)$ and subtracting their total number from Z, one readily arrives at the formula for the activation factor, i.e. the effective fraction of the active electrons $Z_{\rm eff}/Z$ as a function of $T$:
\be
f(T)={Z_{\rm eff}(T) \over Z}= 1 - \int_0^{x(T)} \, \left [ {\chi(x) \over x} - {T \over T_0} \right ]^{3/2} \, x^2 \, dx~,
\label{zeff}
\ee
where $\chi(x)$ is the Thomas-Fermi function, well known and tabulated, of the scaling variable $x = 2 (4/3\pi)^{2/3} m \alpha Z^{1/3}$,  the energy scale $T_0$ is given by
\be
T_0=2 \, \left ( {4 \over 3 \pi} \right )^{2/3} \, m \, \alpha^2 \, Z^{4/3} \approx  30.8 \, Z^{4/3}\, {\rm eV}~,
\label{t0}
\ee
and, finally, $x_0(T)$ is the point where the integrand becomes zero, i.e. corresponding to the radius beyond which all the electrons are active at the given energy $T$. The energy scale $T_0$ in germanium (Z=32) evaluates to  $T_0 \approx 3.1$\,keV. The Thomas-Fermi activation factor for germanium calculated from the formula (\ref{zeff}) is shown by the dashed line in the plot of Fig.~2. One can see that in the shown energy range it reasonably approximates the stepping behavior of the atomic orbitals. The discussed statistical model is known to approximate the average bulk properties of the atomic electrons with a relative accuracy $O(Z^{-2/3})$ and as long as the essential distances $r$ satisfy the condition $Z^{-1} \ll m \alpha r \ll 1$, which condition in terms of the scaling variable $x$ reads as $Z^{-2/3} \ll x \ll Z^{1/3}$. In terms of the formula (\ref{zeff}) for the number of active electrons, the lower bound on the applicability of the model is formally broken at $T \sim Z^{2/3} T_0$, i.e. at the energy scale of the inner atomic shells. However the effect of the deactivation of the inner electrons is small, of order $Z^{-1}$ in comparison with the total number $Z$ of the electrons. On the other hand, at low $T$, including the most interesting region of $T \sim T_0$, the integral in Eq.(\ref{zeff}) is determined by the range of $x$ of order one, where the model treatment is reasonably justified.\\
\begin{figure}[ht]
\begin{center}
 \leavevmode
    \epsfxsize=12cm
    \epsfbox{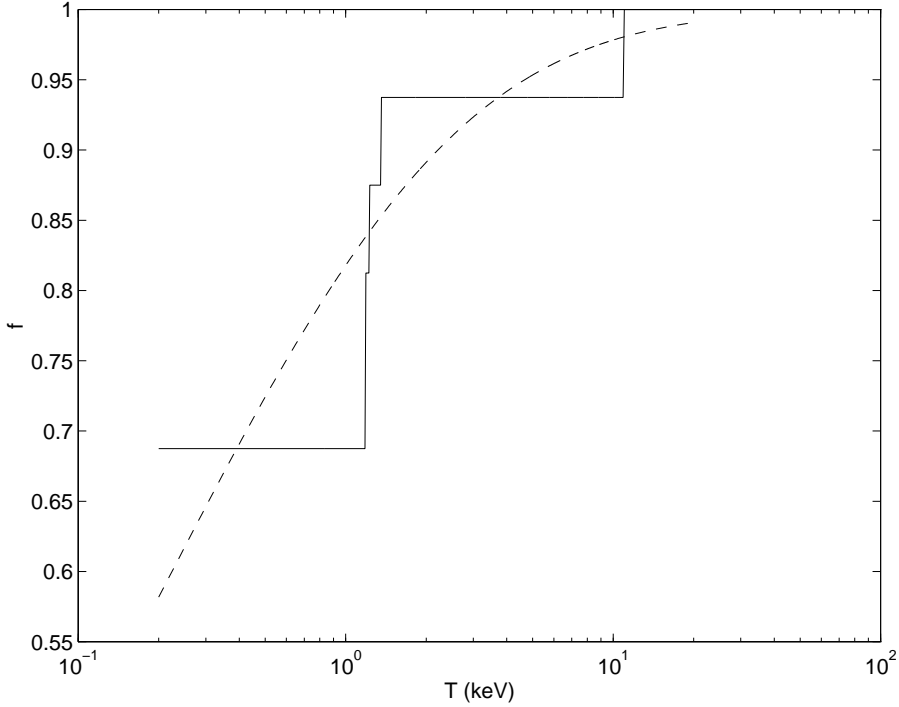}
   \caption{The activation factor $f$ for germanium in the stepping approximation with the actual energies of the orbitals (solid line) and its interpolation in the Thomas-Fermi model (dashed). }
\end{center}
\end{figure}

In order to apply the same model for an estimate of the correlation effect we replace in the estimated correlation contribution to the magnetic neutrino scattering in Eq.(\ref{dic}) the factor $|\psi(0)|^2$ by the total density of the electrons that an active electron `sees' at its location in the atom. Then the resulting correction to the integral $I_1$ for an atom can be written in terms of the density $n_a(r)$ of the active electrons and the total density $n(r)$ of the electrons in the atom:
\be
I_{1c} = -{\pi \, \alpha \over 2 \, m \, T^3} \, \int \, n_a(r) \, n(r) \, d^3r.
\label{ica}
\ee
It should be pointed out that the numerical coefficient in this expression contains a factor of 1/8 as compared to Eq.(\ref{dic}). This is because of a factor of 1/4 corresponding to the statistical weight of the spin-singlet state and an extra 1/2 compensating for the double counting of electrons in the pairs.

One can write the correction described by Eq.(\ref{ica}) in terms of a correction to the activation factor in the Thomas-Fermi model as
\begin{eqnarray}
&&f_c(T) \equiv T \, {I_{1c} \over Z}= \nonumber \\
&&- \left ( {T_1 \over T} \right )^2 \, \left \{ \int_{x(T)}^\infty \,  \chi^3(x) \, {d x \over x} + \int_0^{x(T)} \, \chi^{3/2}(x) \, \left [ \chi^{3/2}(x) - \left ( \chi(x) - { T \over T_0} \, x \right )^{3/2} \right ] \, { dx \over x} \, \right \}~,
\label{dfc}
\end{eqnarray}
where the correlation energy scale $T_1$ is given by
\be
T_1= {\sqrt{2} \over 3 \pi} \, m \, \alpha^2 \, Z \approx 4.1 \, Z \, {\rm eV}
\label{t1}
\ee
and evaluates to about 131\,eV in germanium. The plot of the estimated correlation correction in germanium is shown in Fig.~3. One can readily see that this correction is below 1.5\% at $T \approx 0.3$\,keV and rapidly decreases at higher energy transfer. Clearly, this estimate refers only to the magnetic part of the scattering, while for the weak part we find no correlation effect in the considered order due to the cancellation found in Eq.(\ref{dic}). We thus conclude that in the range of values of $T$ above a few hundred eV the correlation effect can be safely neglected for both contributions to the neutrino scattering on germanium.\\
\begin{figure}[ht]
\begin{center}
 \leavevmode
    \epsfxsize=12cm
    \epsfbox{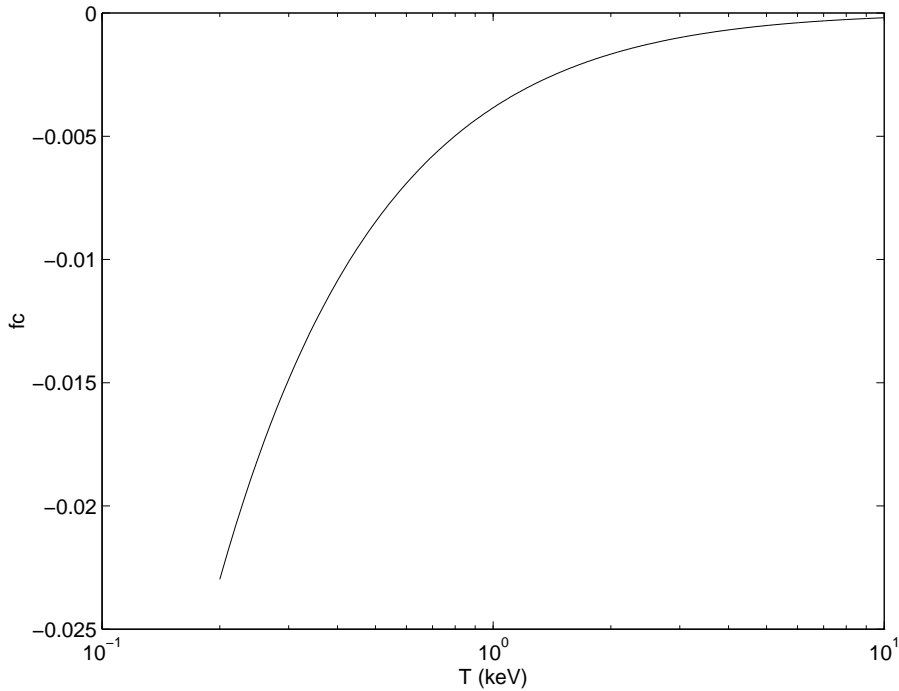}
   \caption{The correlation correction $f_c$ to the activation factor $f$ for germanium in the  Thomas-Fermi model. }
\end{center}
\end{figure}

\section{Summary}
We have considered the scattering of neutrinos on electrons bound
in atoms. Our main finding is that the differential over the
energy transfer cross section given by the free-electron formulas
(\ref{fe}) and (\ref{sew}) and the stepping behavior of the
activation factor given by Eq.(\ref{step}) provides a very
accurate description of the neutrino-impact ionization of a
complex atom, such as germanium, down to quite low energy
transfer. The deviation from this approximation due to the onset
of the ionization near the threshold is less than 5\% (of the
height of the step) for the $K$ electrons, if one applies the
analytical behavior of this onset that we find for the ground
state of a hydrogen-like ion. We also find that the free-electron
expressions for the cross section are not affected by the atomic
binding effects in the semiclassical limit and for independent
electrons. For this reason we expect that the deviation of the
actual onset from a step function at the threshold for ionization
of higher atomic orbitals is even smaller than for the ground
state, since the motion in the higher states is closer to the
semiclassical limit.  The approximation of independent electrons
lacks an account for the two-electron correlations arising from
the Coulomb interaction between the electrons in the atom. We
estimate this effect in the large $Z$ limit using the Thomas-Fermi
model and argue that the effect of the correlations is small in
germanium for the values of the energy transfer above $0.2 \div
0.3$\,keV. We thus argue that for practical applications, i.e. for
the analysis of data of the searches for NMM one can safely apply
the free-electron formulas and the stepping approximation at the
energy transfer down to this range.

\begin{acknowledgments}
We thank A.S. Starostin and Yu.V. Popov for useful and stimulating
discussions. The work of MBV is supported in part by the DOE grant
DE-FG02-94ER40823.
\end{acknowledgments}

\appendix
\section{Sum rules}
\label{A}
We consider here the general sum rules for the dynamical structure
factor $S(T,q^2)$, which stem from the analyticity of the
density-density Green's function $F(T,q^2)$ at a fixed $T$ and
complex $q^2$ and also from its asymptotic behavior at large large
$|q^2|$. At a non zero $T$ the dynamical structure function,
defined by Eq.(\ref{dsf}), vanishes at $q^2=0$, due to the
orthogonality of the excited states $|n \rangle$ and the initial
state $|0 \rangle$ in Eq.(\ref{dsf}) since $\rho(0)$ reduces to a
unit operator. For this reason the function $F(T,q^2)$ is real at
$q^2=0$ and thus satisfies in the complex plane the condition
$F(T, z^*)=F^*(T,z)$. At a non zero real $q^2$ the imaginary part
of this function is not vanishing for both positive and {\it
negative $q^2$}, so that it has cuts along the real axis extending
from zero to both infinities~\footnote{It is not clear what
physical meaning can be ascribed in this problem to negative
$q^2$. However a formal analytical continuation to negative $q^2$
exists and results in a cut along the negative real axis. It is
the omission of this cut that resulted in a somewhat incorrect
treatment of the problem in Ref.~\cite{mv}.}. On the other hand,
the asymptotic at large $|q^2|$ behavior of the Green's function
$F(T,q^2)$ is determined by the free-electron formula (\ref{ff}),
since at $|q^2| \to \infty$ any interaction terms can be
neglected. For a scattering on an atom with $Z$ electrons one
finds 
\be \left. F(T,q^2) \right |_{|q^2| \to \infty} \to - { 2 m
\, Z \over q^2}~. 
\label{asymp} 
\ee 
This behavior enables one to
write a dispersion relation for the Green's function with no
subtractions: 
\be F(T,q^2)= {1 \over \pi} \,
\int_{-\infty}^{\infty} \, {{\rm Im} F(T, Q^2) \over Q^2-q^2 - i
\epsilon} \, dQ^2~. 
\label{disp} 
\ee 
By comparing the dispersion
relation at $q^2 \to \infty$ with the asymptotic behavior in
Eq.(\ref{asymp}) one readily finds the sum rule for an integral
similar to $I_2$, but extended to include also the negative $q^2$:
\be \int_{-\infty}^{\infty} \, S(T, q^2) \, dq^2 = 2m \, Z~,
\label{sr2}
\ee
where the dynamical structure function at negative
$q^2$ is defined by the analytical continuation and
Eq.(\ref{sfrel}), rather than by Eq.(\ref{dsf}).

In order to derive from Eq.(\ref{disp}) a relation for an integral
similar to $I_1$ it is necessary to consider the Green's function
near the origin, i.e. at $q^2 \to 0$. In multi-electron systems
the behavior in this region is generally complicated by the
two-electron correlations. For this reason we limit the
consideration here to the system with just one electron, $Z=1$. In
such a system one has $\rho(\vec q) \to 1$ at $q \to 0$, so that
the Green's function in Eq.(\ref{fdef}) is contributed by only the
initial state $|0 \rangle$:
\be F(T,0)= {1 \over T}~.
\ee
By
comparing this formula with Eq.(\ref{disp}) at $Q^2 \to 0$ one
immediately finds the sum rule
\be \int_{-\infty}^{\infty} \, S(T,
q^2) \, {dq^2 \over q^2} = {1 \over T}~.
\label{sr1}
\ee
It should
be pointed out that unlike the sum rule (\ref{sr2}) this latter
relation is generally invalidated in multi-electron system by the
correlation effects. In fact  an indication of such a difference
in the behavior of the two integrals can be seen in
Eq.(\ref{dic}), where the discussed there correlation effect
vanishes for the integral $I_2$, but not for the $I_1$.

The sum rule (\ref{sr1}) can also be derived from the latter
expression in Eq.(\ref{f1}). Indeed one can rewrite the formula as
\be F(T,q^2) = {1 \over T-{q^2 \over 2m}} + {1 \over T-{q^2 \over
2m}} \, \left \langle 0 \left | {1 \over T -{q^2 \over 2m} - {
(\vec p \cdot \vec q) \over m } - H + E_0 } \, { (\vec p \cdot
\vec q) \over m } \right | 0 \right \rangle~,
\label{f1exp}
\ee
and consider the expansion of the last term in powers of $(\vec p
\cdot \vec q)$. Only the even terms in this expansion are non
vanishing, since the odd terms give zero due to the parity. One
can readily see that in each term in the expansion the pole in
$q^2$ is of a higher order than the power of $q^2$ in the
numerator, so that the imaginary part of each term integrates to
zero in the integral as in Eq.(\ref{sr1}), while the term of the
zeroth order in $(\vec p \cdot \vec q)$ in Eq.(\ref{f1exp}) gives
the sum rule (\ref{sr1}). It is again important here that the
integration runs over all values of $q^2$ i.e. from $-\infty$ to
$+\infty$, since only in this case all the poles of the terms in
the expansion are within the integration range. Any restriction of
the range of integration over $q^2$ may leave some poles out so
that the vanishing of the contribution of all higher terms in the
expansion is generally not guaranteed.

\section{Momentum-transfer integrals for hydrogen-like states}
\label{B}
%
%
Consider the situation when the initial electron occupies the
discrete $nl$ orbital in a Coulomb potential $V(\vec{r})=-\alpha
Z/r$. The dynamical structure factor for this hydrogen-like system
is given by
\be S_{(nl)}(T,q^2)=
\frac{mk}{(2\pi)^3}\frac{1}{2l+1}\sum_{m_l=-l}^l \int
d\Omega_k|\langle\varphi^-_{\vec{k}}|\rho(\vec{q})|\varphi_{nlm_l}\rangle|^2,
\label{sfnl}
\ee
where $\varphi_{nlm_l}$ is the bound-state wave function,
$\varphi^-_{\vec{k}}$ is the outgoing Coulomb wave for the ejected
electron with momentum $\vec{k}$, and
$k=|\vec{k}|=\sqrt{2mT-p_n^2}$, with $p_n=\alpha Zm/n$ being the
electron momentum in the $n$th Bohr orbit. The closed-form
expressions for the bound-free transition matrix elements in
Eq.~(\ref{sfnl}) can be found, for instance, in
Ref.~\cite{belckic81}. In principle, they allow for performing
angular integrations in Eq.~(\ref{sfnl}) analytically. This task,
however, turns out to be formidable for large values of $n$.
Therefore, below we restrict our consideration to the $n=1,2$
states only, which nevertheless is enough for demonstrating the
validity of the semiclassical approach developed in Sec.~\ref{1e}.

Using results of Ref.~\cite{holt69}, we can present the
function~(\ref{sfnl}) when $n=1,2$ as
\begin{eqnarray}
\label{sfnl1} S_{(nl)}(T,q^2)&=&\frac{2^8mp_n^6}{3[1-\exp(-2\pi\eta)]}\frac{q^2f_{nl}(q^2)}{[(q^2-k^2+p_n^2)^2+4p_n^2k^2]^{2n+1}}\nonumber\\
&{}&\times
\exp\left[-2\eta\arctan\left(\frac{2p_nk}{q^2-k^2+p_n^2}\right)\right ] ,
\end{eqnarray}
where the branch of the arctangent function should be used that
lies between 0 and $\pi$, $\eta=\alpha Zm/k$ is the Sommerfeld
parameter, and
\begin{eqnarray}\label{f1s} f_{1s}(q^2)&=&3q^2+k^2+p_1^2,\\
\label{f2s} f_{2s}(q^2)&=&8\left[3q^{10}-(32p_2^2+11k^2)q^8+(82p_2^4+72p_2^2k^2+14k^2)q^6\right.\nonumber\\
&{}&\left.+(20p_2^6-62p_2^4k^2-20p_2^2k^4-6k^6)q^4+(p_2^2+k^2)\right.\nonumber\\
&{}&\left.\times\left(\frac{47}{5}p_2^6-\frac{47}{5}p_2^4k^2-7p_2^2k^4-k^6\right)q^2+(4p_2^2+k^2)(p_2^2+k^2)^4\right],\\
\label{f2p} f_{2p}(q^2)&=&2p_2^2\left[36q^8-48(p_2^2+k^2)q^6+(152p_2^4-48p_2^2k^2-8k^4)q^4+(p_2^2+k^2)\right.\nonumber\\
&{}&\left.\times\left(\frac{1712}{15}p_2^4+\frac{1568}{15}p_2^2k^2+16k^4\right)q^2
+\left(\frac{44}{3}p_2^2+4k^2\right)(p_2^2+k^2)^3\right].
\end{eqnarray}
Insertion of Eq.~(\ref{sfnl1}) into the integrals~(\ref{defi}) and
integration over $q^2$, using the change of variable
$$
\frac{2p_nk}{q^2-k^2+p_n^2}=\tan x
$$
and the standard integrals involving the products of the
exponential function and the powers of sine and cosine functions,
yields
\begin{eqnarray}
\label{mw1s}I_1^{(1s)}(T)=\frac{I_2^{(1s)}(T)}{2mT}&=&\frac{T^{-1}}{1-\exp(-\frac{2\pi}{\sqrt{y_1-1}})}\left\{1-
\exp\left(-\frac{\pi}{\sqrt{y_1-1}}\right)\right.
\nonumber\\&{}&\left.\times\exp\left [ \frac{-2}{\sqrt{y_1-1}}\arctan\left(\frac{y_1-2}{2\sqrt{y_1-1}}\right)\right ]
\left(1-\frac{4}{y_1}+\frac{16}{3y_1^2}\right)\right \},
\end{eqnarray}
\begin{eqnarray}
\label{m2s}I_1^{(2s)}(T)&=&\frac{T^{-1}}{1-\exp(-\frac{4\pi}{\sqrt{y_2-1}})}\left \{ 1-
\exp\left(-\frac{2\pi}{\sqrt{y_2-1}}\right)\right.
\nonumber\\&{}&\left.\times\exp\left [ \frac{-4}{\sqrt{y_2-1}}\arctan\left(\frac{y_2-2}{2\sqrt{y_2-1}}\right)\right ]
\left(1-\frac{8}{y_2}+\frac{80}{3y_2^2}-\frac{448}{15y_2^3}+\frac{1792}{15y_2^4}\right)\right \},\nonumber\\
\end{eqnarray}
\begin{eqnarray}
\label{w2s}I_2^{(2s)}(T)&=&\frac{2m}{1-\exp(-\frac{4\pi}{\sqrt{y_2-1}})}\left \{ 1-
\exp\left(-\frac{2\pi}{\sqrt{y_2-1}}\right)\right.
\nonumber\\&{}&\left.\times\exp\left [ \frac{-4}{\sqrt{y_2-1}}\arctan\left(\frac{y_2-2}{2\sqrt{y_2-1}}\right)\right]
\left(1-\frac{8}{y_2}+\frac{80}{3y_2^2}-\frac{448}{15y_2^3}+\frac{1024}{15y_2^4}\right)\right \},\nonumber\\
\end{eqnarray}
\begin{eqnarray}
\label{m2p}I_1^{(2p)}(T)&=&\frac{T^{-1}}{1-\exp(-\frac{4\pi}{\sqrt{y_2-1}})}\left \{ 1-
\exp\left(-\frac{2\pi}{\sqrt{y_2-1}}\right)\right.
\nonumber\\&{}&\left.\times\exp\left [ \frac{-4}{\sqrt{y_2-1}}\arctan\left(\frac{y_2-2}{2\sqrt{y_2-1}}\right)\right ]
\left(1-\frac{8}{y_2}+\frac{80}{3y_2^2}-\frac{704}{15y_2^3}+\frac{3328}{45y_2^4}\right)\right \},\nonumber\\
\end{eqnarray}
\begin{eqnarray}
\label{w2p}I_2^{(2p)}(T)&=&\frac{2m}{1-\exp(-\frac{4\pi}{\sqrt{y_2-1}})}\left \{ 1-
\exp\left(-\frac{2\pi}{\sqrt{y_2-1}}\right)\right.
\nonumber\\&{}&\left.\times\exp\left [ \frac{-4}{\sqrt{y_2-1}}\arctan\left(\frac{y_2-2}{2\sqrt{y_2-1}}\right)\right ]
\left(1-\frac{8}{y_2}+\frac{80}{3y_2^2}-\frac{704}{15y_2^3}+\frac{512}{15y_2^4}\right)\right \} ,\nonumber\\
\end{eqnarray}
where $y_n=2mT/p_n^2\equiv T/|E_n|$. The largest deviations of
these integrals from the free-electron analogs~(\ref{intf}) occur
at the ionization threshold $T=|E_n|$. The corresponding relative
values in this specific case are
$$
\frac{I^{(1s)}_1}{I^{(FE)}_1}=\frac{I^{(1s)}_2}{I^{(FE)}_2}=1-\frac{7}{3}e^{-4}=0.9572635093,
$$
$$
\frac{I^{(2s)}_1}{I^{(FE)}_1}=1-\frac{1639}{15}e^{-8}=0.9633451168,
\qquad
\frac{I^{(2s)}_2}{I^{(FE)}_2}=1-\frac{871}{15}e^{-8}=0.9805208034,
$$
$$
\frac{I^{(2p)}_1}{I^{(FE)}_1}=1-\frac{2101}{45}e^{-8}=0.9843376226,
\qquad
\frac{I^{(2p)}_2}{I^{(FE)}_2}=1-\frac{103}{15}e^{-8}=0.9976964900.
$$
The above results indicate a clear tendency: the larger $n$ and
$l$, the closer $I_1^{(nl)}$ and $I_2^{(nl)}$ are to the free-electron
values. The departure from the free-electron behavior does not
exceed several percent at most. These observations provide a solid
base for the semiclassical approach of Sec.~\ref{1e}.

\end{document}